\documentclass[journal,draftclsnofoot,onecolumn,12pt]{IEEEtran}
\usepackage[letterpaper,margin=1.6cm,footskip=0.4cm]{geometry}
\usepackage{graphicx}
\usepackage{epsfig}
\usepackage{algorithm}
\usepackage{algorithmic}
\usepackage{ifmtarg}
\usepackage{cite}
\usepackage{amssymb}
\usepackage{amsthm}
\usepackage{amsmath}
\usepackage{color}
\usepackage{bbm}
\usepackage{cite}
\usepackage{epstopdf}
\usepackage[table,xcdraw]{xcolor}

\newtheorem{definition}{Definition}

\usepackage{subcaption}

\newcommand\blfootnote[1]{%
  \begingroup
  \renewcommand\thefootnote{}\footnote{#1}%
  \addtocounter{footnote}{-1}%
  \endgroup
}

\begin{document}
\title{Spatial Firewalls: Quarantining Malware Epidemics in Large Scale Massive Wireless Networks}

\author{ \IEEEauthorblockN{Hesham ElSawy, {\em Senior Member, IEEE}, Mustafa A. Kishk, {\em Member, IEEE}, \\ and Mohamed-Slim Alouini, {\em  Fellow, IEEE} }}
\maketitle

\vspace{-6mm}

\begin{abstract}

Billions of wireless devices are foreseen to participate in big data aggregation and smart automation in order to interface the cyber and physical worlds. Such large-scale ultra-dense wireless connectivity is vulnerable to malicious software (malware) epidemics. Malware worms can exploit multi-hop wireless connectivity to stealthily diffuse throughout the wireless network without being noticed to security servers at the core network. Compromised devices can then be used by adversaries to remotely launch cyber attacks that cause large-scale critical physical damage and threaten public safety. This article overviews the types, threats, and propagation models for malware epidemics in large-scale wireless networks (LSWN). Then, the article proposes a novel and cost efficient countermeasure against malware epidemics in LSWN, denoted as spatial firewalls. It is shown that equipping a strategically selected small portion (i.e., less than $10\%$) of the devices with state-of-the-art security mechanisms is sufficient to create spatially secured zones that quarantine malware epidemics. Quarantined infected devices are then cured by on-demand localized software patching. To this end, several firewall deployment strategies are discussed and compared.


\end{abstract}


\section{Introduction}

\blfootnote{H. ElSawy is with King Fahd University of Petroleum and Minerals (KFUPM), Dhahran, Eastern Province, Saudi Arabia. email: {hesham.elsawy@kfupm.edu.sa.} \\ 	
 M.A. Kishk and M.-S. Alouini are with King Abdullah University of Science and Technology (KAUST), Thuwal, Makkah Province, Saudi Arabia, email: \{mustafa.kishk, slim.alouini\}@kaust.edu.sa. \\
This work is funded in part by the deanship of scientific research (DSR), at King Fahd University of Petroleum and Minerals (KFUPM), under research grant no. DF191052.}

The imminent era of smart world relies on large-scale massive wireless connectivity that interfaces the physical and cyber worlds. The surging Internet of Things (IoT) and cyber physical systems (CPS) with massive numbers of heterogeneous wireless devices (e.g., sensors, actuators, smart phones, smart appliances, autonomous vehicles, etc.) are examples of such massive large-scale wireless networks (LSWNs). IoT/CPS are foreseen to provide flexible platforms for big data aggregation and/or smart automation to almost every aspect in our lives~\cite{CPS_Sec_Survey}. 
For instance, intelligent transportation systems with  connected/autonomous vehicles exploit wireless connectivity to improve road safety and reduce traffic congestion. Smart power grids utilize wireless connectivity for data communications and smart control (e.g., smart meters and field devices) in order to enhance energy generation and  distribution. Large scale massive connectivity is also a foundational building block for process automation in the next industrial revolution (i.e., industry 4.0). In addition to the aforementioned examples, large-scale massive wireless connectivity can bring unlimited potentials to many other verticals such as health care, public safety, agriculture, retail, etc. 

On the downside, massive LSWNs bring a multitude of new and challenging security threats~\cite{CPS_Sec_Survey}. Particularly, many of the wireless devices in the IoT/CPS are installed and controlled via consumers with limited security background. 
{The high competition between IoT/CPS manufacturers overlooks cybersecurity aspects to  reduce costs and keep-up with the rapid proliferation of IoT/CPS}. Many IoT/CPS devices are too constrained, in terms of computational power, energy, and storage, to implement and continuously execute sophisticated defense mechanisms~\cite{jumping, Poisson_Games}. Such lack of security oriented network administration and per-device defense mechanisms opens several loopholes for adversaries to infiltrate malicious software, or shortly malware, to the network. 

{Conventionally, malware programs are designed for variety of criminal and hostile activities such as spying (spyware), threatening for monetary benefit (ransomware), and/or controlling large population of devices (botnets). In  IoT/CPS networks, malware hostile activities naturally extend to physical threats. In smart vehicles, adversaries can control the vehicle through telematics unit, which introduce the risk of physical denial of service (DoS) (e.g., stop the engine and lock doors/windows) as well as deliberate collisions. In power grids, adversaries can compromise field devices (e.g., switch gears and circuit breakers) to sabotage equipment, disrupt power distribution, and cause major blackouts. In medical care systems, adversaries can inject false prescriptions and manipulate wearable drug infusion devices. In industrial environments, adversaries can halt/manipulate ongoing production lines,  delete customized machine setting, or even damage products and injure workers. There could be also generic attacks for LSWN such as network jamming and colluded eavesdropping. Note that network-jamming attacks may disrupt the entire network connectivity via overwhelming interference and shorten the network lifetime by depleting compromised devices batteries. Using colluded eavesdropping, adversaries can reveal and misuse private data. }

Exploiting the massive spatial density in LSWN and multi-hop wireless connectivity (e.g., machine-to-machine communications), the malware can stealthily propagate from one device to another and form an {\em epidemic outbreak} without being noticed by the security administration at the core network \cite{jumping, Vehicular_epidemics}. Even worse, the emerging beyond 5G technologies (e.g., non-orthogonal multiple access (NOMA) and ultra-reliable low latency communications (URLLC)), that are meant to enhance information dissemination, will also accelerate epidemic outbreak throughout the network. An epidemic outbreak of a malware enables the adversities to control a large population of devices and launch large-scale cyberphiscal attacks, which may lead to catastrophic consequences in IoT/CPS. From the propagation point of view, malware can be classified into the following categories 

\begin{itemize}
	
	\item \textbf{Trojans:} Malware hiding in legitimate programs but intended to infect a target system and open backdoors for future intrusion. 
	
	\item \textbf{Virus:} Self-replicating malware designated to infect and corrupt the operation of a target system. Viruses propagates via host executable files.  
	
	\item \textbf{Worm:} Self-replicating malware intended to spread and infect all the devices in a network. A worm is a stand-alone software that automatically propagate from one device to another. 
	
\end{itemize}


Human interventions, such as manual  attacks by adversaries or infected file exchange/execution by legitimate users,\footnote{Legitimate users can be incentivised to install free programs that contain malware.} are required for Trojans and viruses to spread in a network.  On the other hand, worm malware automatically identifies network vulnerabilities to diffuse and compromise new targets. Exploiting the dense network deployment, the broadcast nature of the wireless channel, the one-to-many communication schemes (e.g, NOMA), and high-reliability-low-latency communications (e.g., URLLC), worm malware can quickly and covertly spread in the wireless network and form an epidemic outbreak. This makes worm epidemics the hardest to decelerate their diffusion and/or quarantine their infection. Hence, worm malware is among the highest security threats for large-scale wireless networks~\cite{Nature_worm}. 

\section{Securing large-scale IoT/CPS}
The aforementioned threatening physical consequences of malware intrusion in CPS/IoT call out for robust security countermeasure and defense mechanisms. Securing the devices against manipulation and intrusion is the first line of defense for IoT/CPS networks. Such defense mechanisms could be embedded in hardware (e.g., trusted platforms), software (e.g., anti-malware programs), communication protocols  (e.g., encryption \&  authentication), and/or device operation (scheduled attestation/patching). However, the strict cost, energy, and computational power constraints of devices in many IoT/CPS applications limit the implementation of sophisticated defense mechanisms to all devices. {In this regards, \cite{Poisson_Games} proposes a Poisson game to distributively decide on which devices to adopt an anti-malware such that an epidemic outbreak is prevented. However, the proposed mechanism in \cite{Poisson_Games} is based on a fully mixed epidemic model,\footnote{A fully mixed epidemic model assumes that an infection (e.g., malware) can be directly transmitted from any node in the network to any other node in the network.} which overlooks the spatial topology and limited wireless communication range in LSWN. 
	
{Another effective, yet simple, defense mechanism is to perform scheduled (i.e., periodic) software attestation/patching for IoT/CPS devices to ensure configurations integrity and wipeout potential malicious software~\cite{Botnet, seda}. However, to thwart epidemic diffusion, the treatment rate should be faster than the epidemic infection rate. Being oblivious to the device status, unnecessary disruption for the IoT/CPS operation may occur due to attesting/patching healthy devices~\cite{Botnet}. Furthermore, a device that is compromised shortly after being patched/attested, may have enough time to be exploited to launch versatile malicious attacks. Such problems are more acute when employing  wireless technologies such as NOMA and URLLC due to the accelerated epidemic infection rate.  }

{To efficiently balance the tradeoff between cybersecurity, anti-malware license cost, and devices hardware complexity, this paper proposes a novel ubiquitous security countermeasure, denoted as ``\textbf{\em spatial firewalls}'', which is meant to detect, spatially quarantine, and report malware infections in LSWN. The proposed spatial firewalls countermeasure is detailed in Section III. Section~IV highlights the mathematical propagation models for malware in LSWN, which are necessary to design and assess cybersecurity countermeasures. Section~V showcases and assesses the spatial firewalls solution before the paper is concluded in Section~VI. }

\begin{figure*}[t!]
	\centering
	{\includegraphics[width= 1 \textwidth]{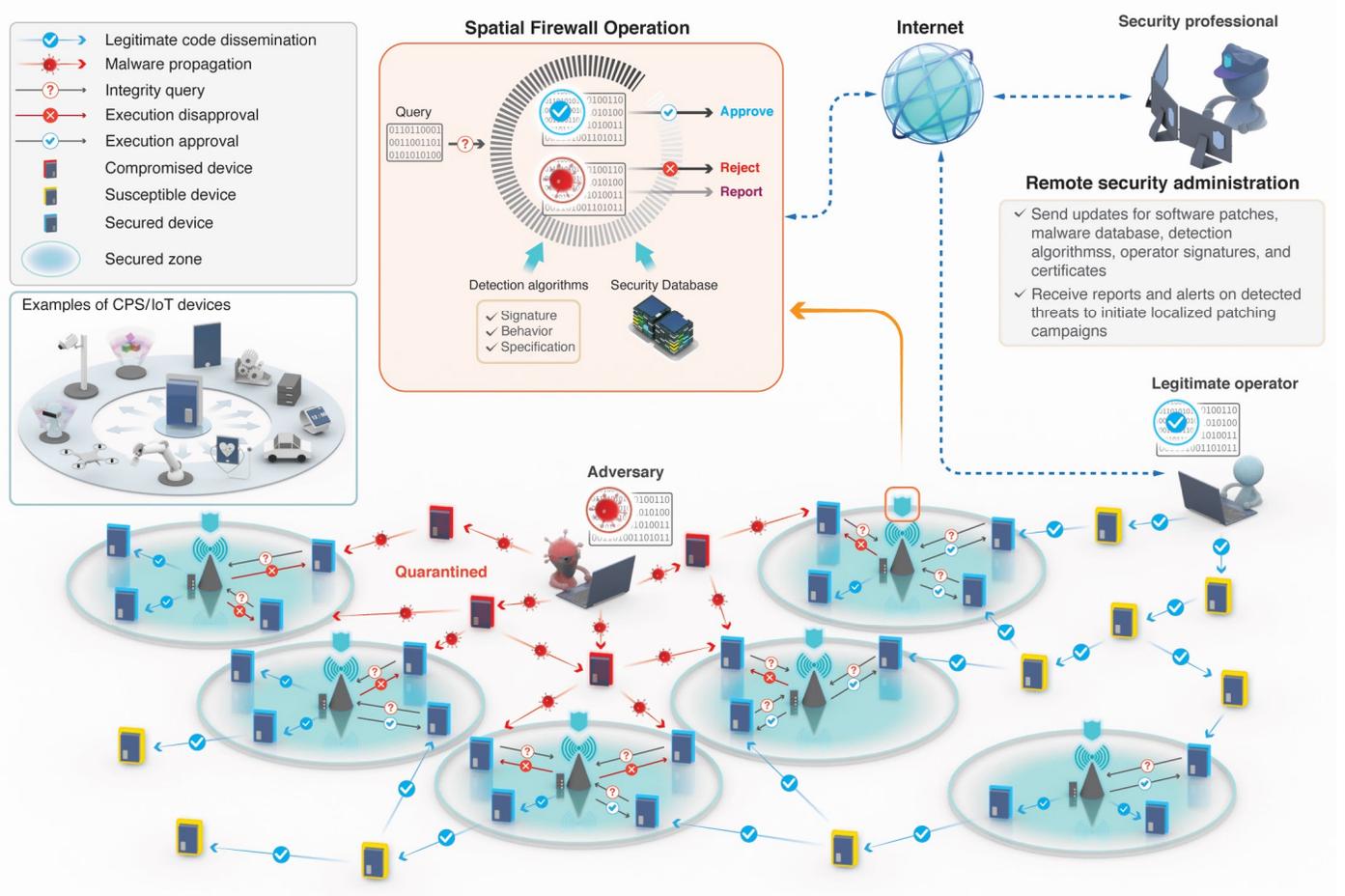}}	
	\caption{Illustration of an edge-computing enabled spatial firewalls operation in massive LSWN to quarantine malware diffusion. The secured zones succeed to spatially quarantine the malware infection within a confined area with limited number of devices.}
	\label{fig:SF0}
\end{figure*}

\section{The spatial firewalls solution}

To secure massive LSWN against malware epidemics, this paper proposes to implement ``\textbf{\em spatial firewalls}'', which are defined as follows:

\begin{definition}
	\textbf{Spatial Firewalls} are wireless devices, with sufficient computational capabilities, energy resources, and memory, to store, execute, and frequently update anti-malware and intrusion detection programs (e.g., edge computing devices, access points, junction nodes). These devices are deployed at critical locations within a LSWN to enforce secured zones in order to quarantine any emerging malware epidemic and {thwart} its outbreak.  
\end{definition}

{ Fig.~\ref{fig:SF0} illustrates an edge-computing enabled spatial firewall operation in massive LSWNs, where firewalls are implemented at base stations. Devices that are adjacent to a firewall inquire about codes (e.g., software updates, system configurations, or control commands) received from their wireless interface. For each validation inquiry, the firewall ensures that the code is malware free (e.g., using signature-based or anomaly-based detection) and verifies its integrity (e.g., using operator digital signature and certificates)~\cite{seda}. Only validated codes are approved for execution and/or dissemination to other devices. If a threat is detected, the firewall disapproves the code execution/dissemination and reports the incident to the security administration. Hence, each firewall creates a {\em secured zone}, determined by its wireless range, for adjacent devices and thwarts their infection (i.e., analogous to herd immunity effect). Thus, firewalls enable an exclusive dissemination of legitimate codes through the network.\footnote{Different from legacy IT systems, IoT/CPS networks have no clear boundary between secured and unsecured (i.e., public) domains. Hence, the proposed spatial firewalls enforce the concept of secured zones within the wireless network.} In summary, spatial firewalls are meant to i) create {\em spatially secured zones} within the wireless network, ii) thwart the dissemination of malware (i.e., ensure that malware can only infiltrate to limited number, by design, of devices regardless of the infection/treatment rates), and iii) initiate {\em on-demand software attestation/patching} for infected devices. }

{ It is worth noting that extending the firewall code validation scheme for non-adjacent devices through multi-hop wireless inquiries may impose overwhelming signaling overhead and large control latency due to the massive and wide-scale deployment of devices. To balance the security, signaling overhead, and control latency of the network, the code-validation role of the firewall is limited to adjacent devices. As such, the design objective is to have sufficient numbers of efficiently located firewalls to satisfy the security requirements of the network, which may tolerate interim infection of some devices (i.e., until being detected and patched).}

{As compared to scheduled patching/software attestation~\cite{Botnet, seda}, spatial firewalls provide ubiquitous security countermeasure that reacts to attacks by i) thwarting malware diffusion, ii) localizing infected (i.e., quarantined) regions,\footnote{The location of quarantined regions can be inferred from the reporting firewalls identities and known locations.}  and iii) initiating localized software attestation/patching campaigns. Due to the overwhelming overhead and time required for brute-force software attestation/patching of all devices, localizing infected regions is necessary in massive LSWN.} 

{\subsection{Practical Implementation Challenges}
	To conceptualize and materialize the spatial firewalls cybersecurity countermeasure, several practical challenges should be taken into consideration, which include 
	\begin{itemize}
	\item \textbf{Firewall deployment:} The main technical challenge is to determine the number of firewalls and their spatial locations that guarantee spatially quarantined malware. 
	\item \textbf{Techno-economic aspects:} The firewall deployment should account for the trade-off between network security, involved capital-expenditure (CAPEX) to deploy firewalls, and  operational-expenditure (OPEX) to maintain their up-to-date anti-malware programs and licenses. In case of multiple IoT/CPS operators/owners, an agreement for bearing such CAPEX and OPEX is required.
	\item \textbf{Operator privacy:} In case of multiple operators, code verification and software attestation schemes should keep the specific information and configurations of the operators hidden from each other. 
	\item \textbf{Devices heterogeneity:} For universal utilization of spatial firewalls, unified/standardized signaling protocols for different IoT/CPS devices using different wireless interfaces are required. Such signaling protocols should define the method for firewall discovery, association, and templates for code validation inquiries/responses.
    \item \textbf{Signaling overhead:} Efficient signaling schemes are required for code verification to impose minimum disruption to the existing data traffic. For instance, encrypted digital signature for firewalls should be developed to eliminate redundant validation of the same code across different firewalls.    
	\item \textbf{Latency:} The code validation and execution should occur within the tolerable latency defined by data communications and/or control applications. This may require developing security-aware traffic prioritization schemes that expedite firewalls related signaling.    
	\item \textbf{Devices Mobility:} Malware may get through the secured-zones by means of physical mobility of infected devices. Hence, the spatial firewalls should be able to detect and cure infected mobile devices passing through secured zones.
\end{itemize}
This paper focuses on the spatial firewall deployment that ensures spatially quarantined malware. Other challenges are left for future extensions. In particular, we aim to provide guidelines to design and assess the impact of implementing spatial firewalls. To minimize the associated CAPEX and OPEX, the objective is to find the minimum number of firewalls that enforce an epidemic free network operation. For this purpose, we first present the underlying mathematical models for malware propagation that are used to formulate and solve the spatial firewall design problem. Then, several spatial deployment strategies for firewalls are discussed and compared. }
\section{Propagation Models for Malware Epidemics}\label{sec:models}
Mathematical models that characterize propagation of infection in large populations are used to predict epidemic outcomes and design defense mechanisms. In IT systems, the population represents a network of connected devices. {The network topology is mathematically  described by a graph $G=\{V,E\}$, where $V$ is the set of vertices (i.e., devices) and $E$ is the set of edges connecting the vertices (see Fig.~\ref{fig:SF1}). {In the context of wireless networks, an edge between two devices implies that they are within the communication range of each other. The communication range is usually defined by a minimum required signal-to-interference-plus-noise-ratio (SINR)~\cite{SINR_graphs}}. Graphs that account for the random spatial locations of the devices and their wireless communications ranges are denoted as random geometric graphs (RGGs). }

  An epidemic is considered as a process on the graph, where each device (i.e., vertex) can transition between different states such as susceptible (S) (i.e., healthy but can be infected), infected (I) (i.e., compromised via a malware), and recovered (R) (i.e., malware is detected and removed). A malware worm can only infiltrate from an infected device to a susceptible device if the two devices are directly connected via an edge. Once a new device gets compromised by a malware, it becomes an infection threat to its directly connected neighbors and so on.

The dynamics of epidemic infection/treatment are fully characterized via time domain models. Due to several factors (e.g., medium access control protocols and per-link transmission rate), the time taken for malware worm to infiltrate from an infected to susceptible device is random. Meanwhile, depending on the malware detection and treatment technique, the time a device stays infected is also random. On average, propagation and recovery rates (or probability per unit time) can be characterized, which are then used to construct a system of differential equations that fully describes the temporal evolution of an epidemic. Resorting to the fact that the total number of devices in the network remain fixed (i.e., the total population), such system of equations can be solved to determine the percentiles of devices in each of the S, I, and R states, as function of time. {However, in RRGs, such system of differential equations is not tractable and approximations are always sought~\cite{newman_networks}. }

Instead of full temporal characterization, the final outcome of an epidemic infection can be directly characterized. Such late-time characterization alleviates the mathematical complications (e.g., non-linearity of differential equations) introduced by temporal models. In particular, late-time models characterize the overall epidemic infiltration through the network without any information about the infection/treatment rates. As mentioned earlier, the time taken for a malware worm to propagate from a device to each of its neighbors is random and the time each device remain infected is also random. Hence, even after sufficiently long time, some devices will be cured before infecting some of their neighbors and the malware worm would only infiltrate through a subset of the network. For a given malware worm, the late time model would show all devices that were infected at any point in time regardless if they got cured or not. Such phenomenon of global and time oblivious worm diffusion, through a portion of the network, can be studied via percolation theory, which is defined as follows

\begin{definition}
	\textbf{Percolation Theory} is a well-developed mathematical field that characterizes  global connectivity in random graphs when vertices and their associated edges are removed. Connectivity is characterized by the presence/absence of a {\em giant component}, which is the largest connected sub-graph that spans the horizon. 
\end{definition}

Mapping to the aforementioned epidemic models, percolation theory can be used to characterize late-time epidemic outbreak as follows. Consider the complete network graph and remove all devices, with their associated edges, that are never infected (i.e., their neighbors are cured before malware infiltration). 
After such devices removal, the existence (absence) of a giant component implies an epidemic outbreak (quarantine). In the notation of percolation theory, we have the following definitions

\begin{figure*}[t!]
	\centering
	\begin{subfigure}[t]{0.45\textwidth}
		\centerline{\includegraphics[width=  3.2in, trim={3cm 1cm 2cm 0.8cm},clip]{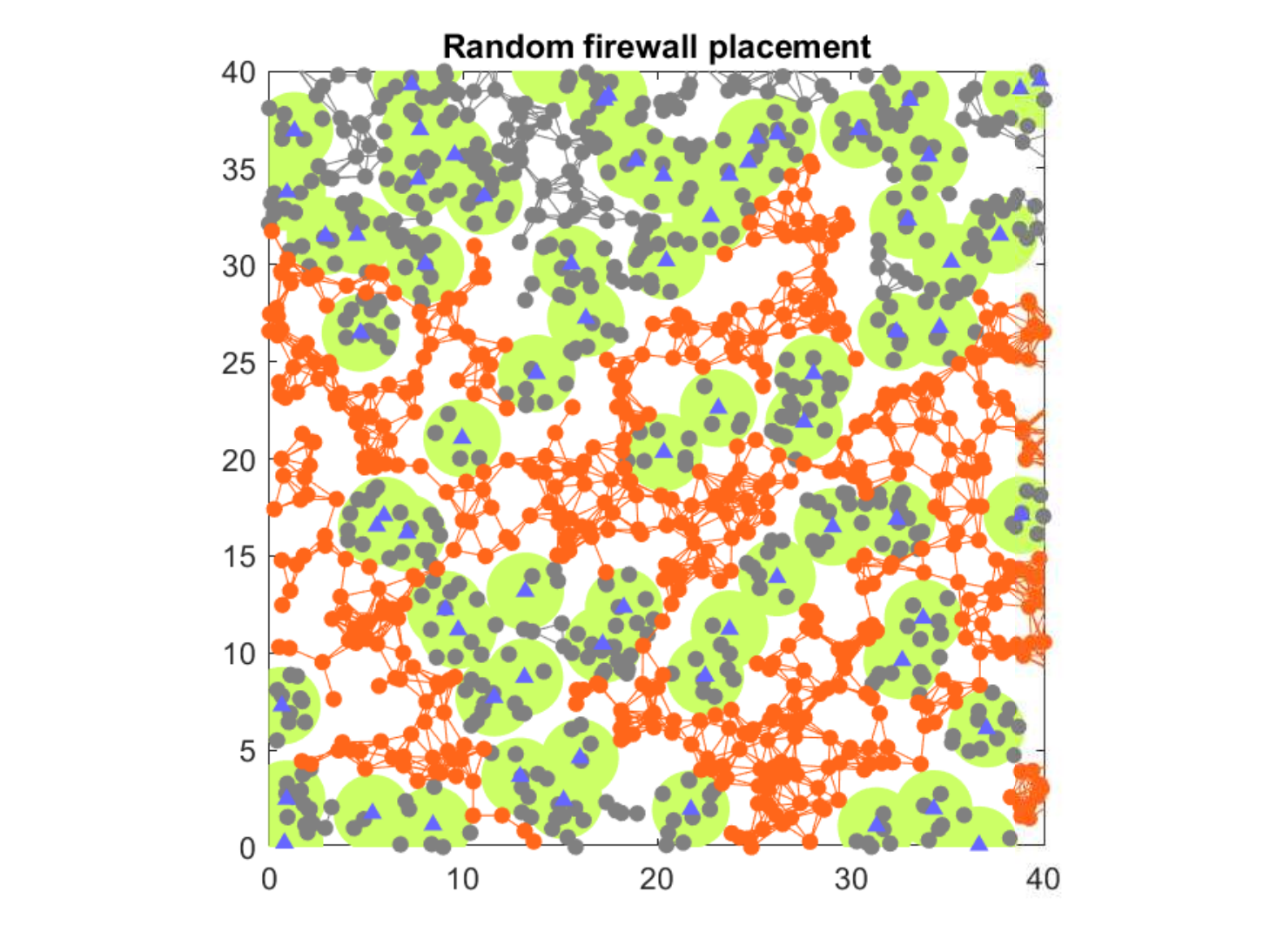}}
		\caption{ Malware epidemic outbreak for Random selection of firewalls \\
			$\quad$}
		\label{fig:random_fire}
	\end{subfigure}
	\begin{subfigure}[t]{0.45\textwidth}
		\centerline{\includegraphics[width=  3.2in, trim={3cm 1cm 2cm 0.8cm},clip]{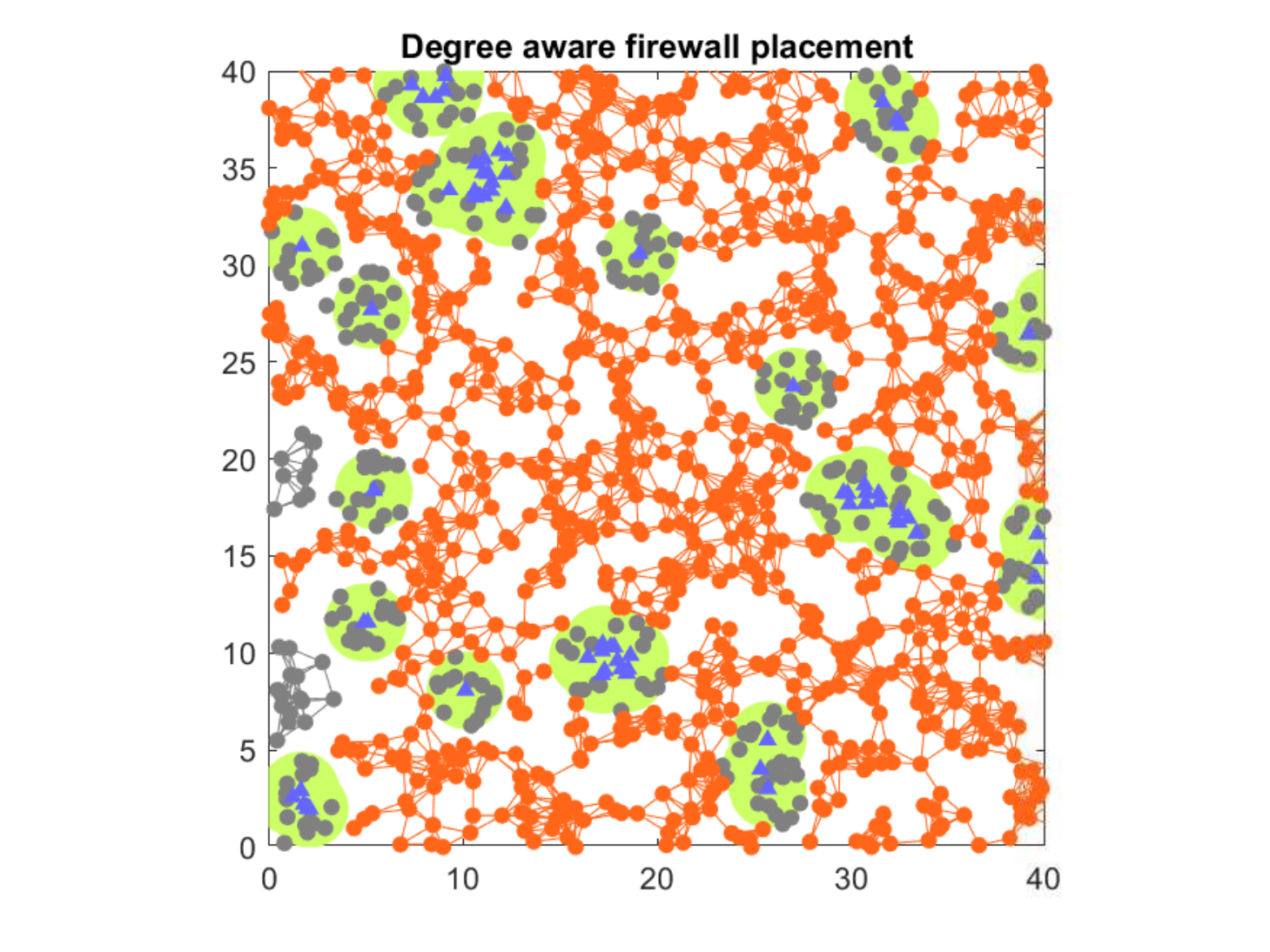}}
		\caption{ Malware epidemic outbreak for degree-aware selection of firewalls}
		\label{fig:degree_fire}
	\end{subfigure}
		\begin{subfigure}[t]{0.45\textwidth}
			\centerline{\includegraphics[width=  3.2 in, trim={3cm 1cm 2cm 0.8cm},clip]{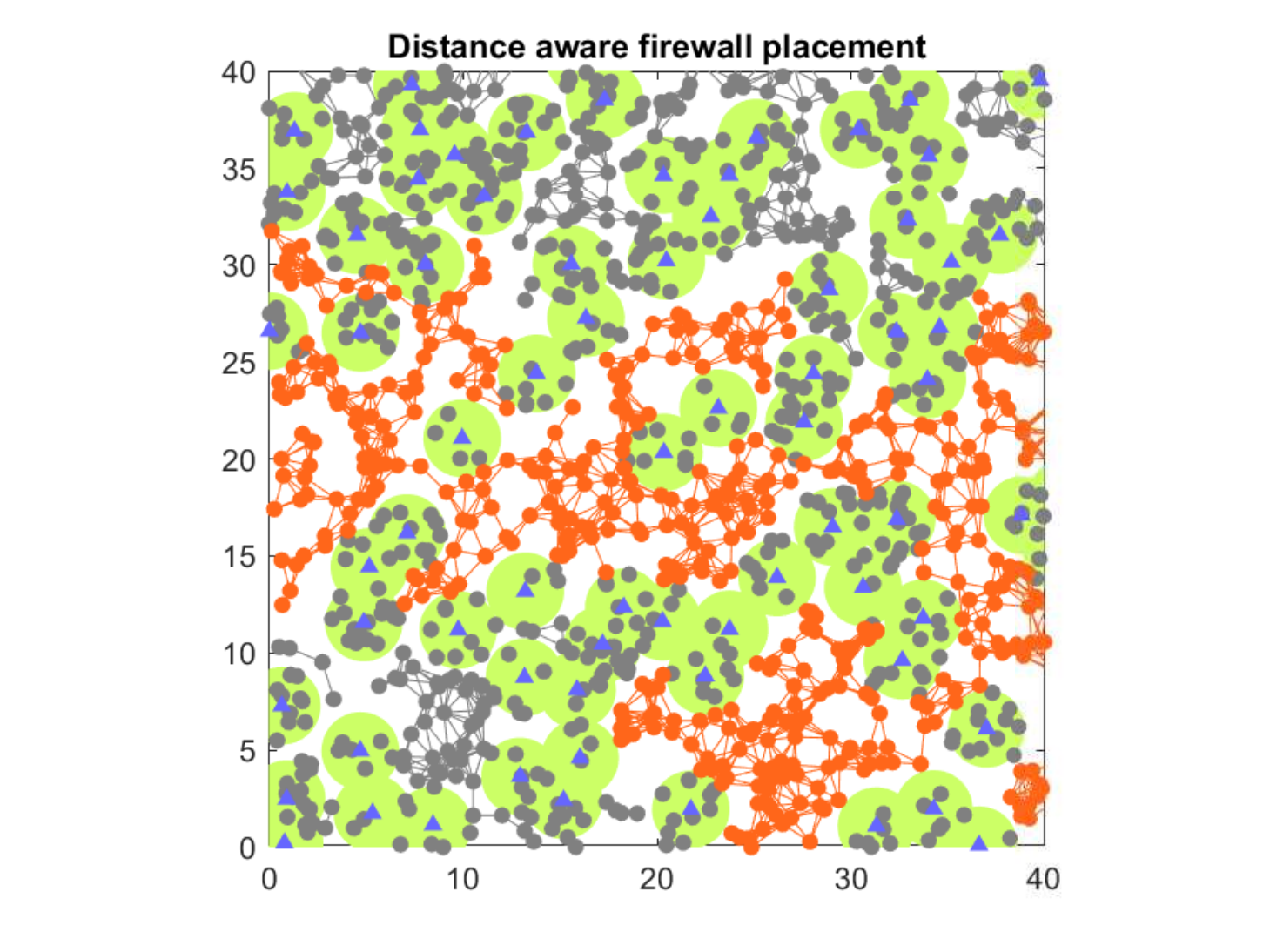}}
			\caption{Malware epidemic outbreak for random selection with minimum DC}
			\label{fig:RDC_fire}
		\end{subfigure}
	\begin{subfigure}[t]{0.45\textwidth}
		\centerline{\includegraphics[width=  3.2 in, trim={3cm 1cm 2cm 0.8cm},clip]{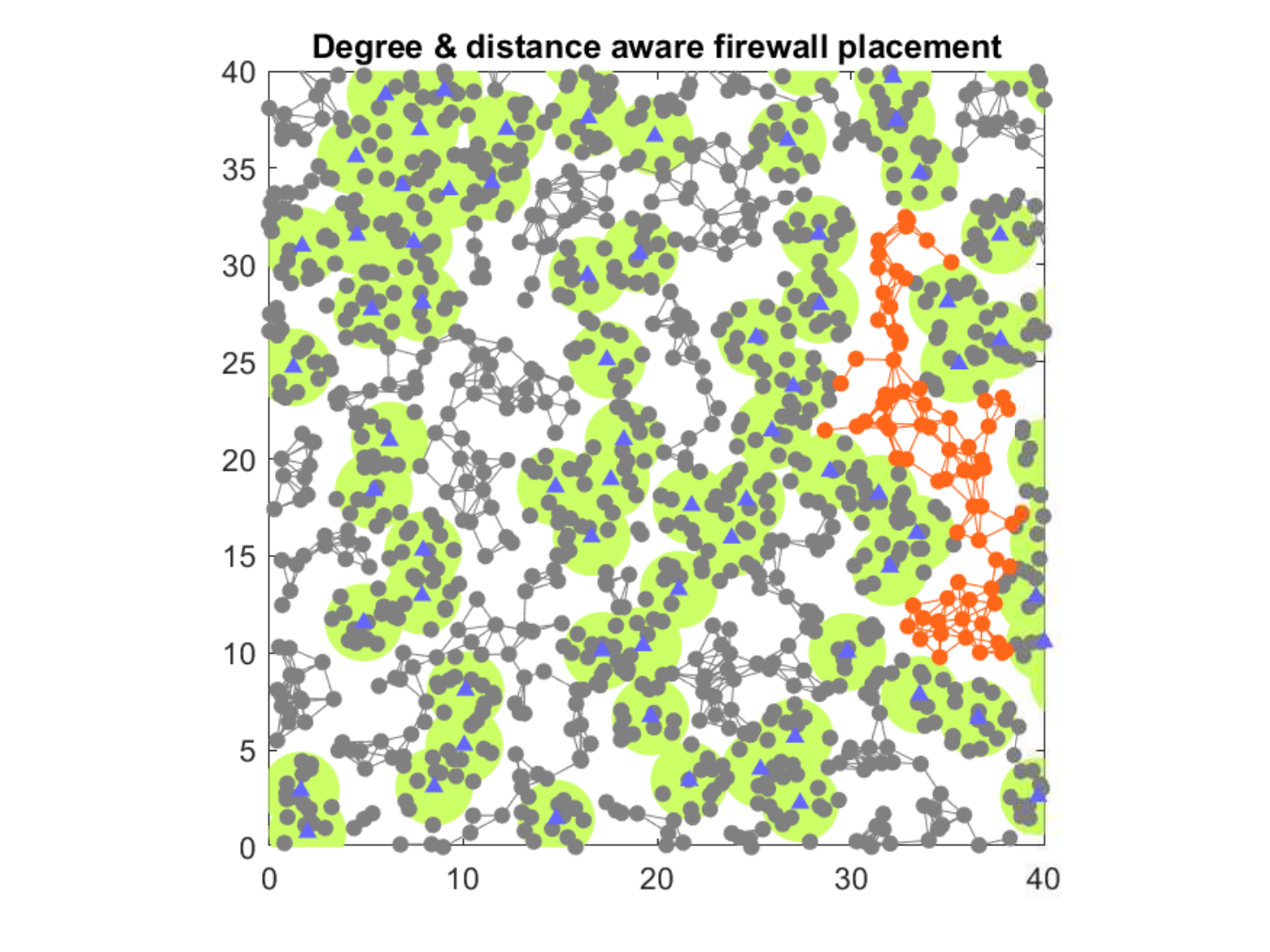}}
		\caption{Quarantined malware for degree-aware selection with minimum DC}
		\label{fig:degree_DC_fire}
	\end{subfigure}
	
	\caption{Epidemic potential diffusion in the same network realization under different firewalls selection strategies for firewalls percentage of $6\%$ of the total network devices. The figure follows the same color code for {\em firewalls}, {\em secured zone}, {\em susceptible}, and {\em compromised} IoT/CPS devices. The links between devices denote potential routes for malware epidemic diffusion, which are obstructed by {\em secured zones}. Only figure (d) eliminates the giant connected component, which implies quarantined epidemic.}
	\label{fig:SF1}
\end{figure*}

\begin{definition}
	\textbf{Supercritical Regime} defines the set of network parameters such that a giant connected component exits. 
\end{definition}

\begin{definition}
	\textbf{Subcritical Regime} defines the set of network parameters such that a giant connected component does not exist. 
\end{definition}

Percolation theory is best suited to develop robust security countermeasures that are independent of the epidemic infection/treatment rates. That is, no matter how fast (slow) is the infection (treatment) rate, a defense mechanism that operates the susceptible  devices in the subcritical regime ensures quarantined epidemic. On the other hand, if the susceptible devices operate in the supercritical regime, then an epidemic outbreak may occur if the treatment rate is not sufficiently faster than the infection rate. In the latter case, the giant component size relative to the network size represents the percentile of infected devices.  

{It is worth noting that percolation theory is a well developed field that has been extensively used in the literature to characterize ad hoc and cognitive network connectivity~\cite{SINR_graphs, per_cog}, secure dissemination of information~\cite{Pinto}, network resilience/reliability~\cite{Cascaded}, and Internet malware epidemics~\cite{per_internet}. However, percolation models are barely used to characterize and develop defense mechanisms for epidemics in LSWN. Note that, in wireless networks, epidemic propagation is highly restricted by the network spatial topology, medium access control, and physical layer properties~\cite{wireless_worm}. Hence, it is important to consider RRG to account for the intrinsic properties of wireless networks~\cite{Pinto,Botnet,SINR_graphs}. In this regards, stochastic geometry (see \cite{survey_h,  tutorial_h, moe_win, survey_martin, martin_book} for tutorials) can be utilized to model the spatial devices locations and construct SINR-aware edges that connect devices that can reliably communicate to each other. Then, exploiting the rich literature on percolation theory, epidemic propagation can be characterized, and defense mechanism can be developed for a variety of IoT/CPS systems and use cases.  }

\section{Spatial Firewalls Design and Assessment}

\subsection{Network Model and Proposed Approach}

{As discussed earlier, due to cost, energy, and computational power constraints, not all IoT/CPS devices can adopt sufficient security mechanisms. Furthermore, state oblivious scheduled software attestation/patching may deteriorate the IoT/CPS functionality due to unnecessary (delayed) patching of healthy (infected) devices. Hence, we propose to strategically select some devices, denoted as spatial firewalls, and equip them with state-of-the-art security mechanisms such as anti-malware and anomaly/intrusion detection programs. As shown in Fig.~\ref{fig:SF0}, each firewall enforces a {\em security zone} of radius $R_f$ meters, where $R_f$ is the communication/detection range of firewalls. Then, all IoT/CPS devices within a secured zone, denoted as {\em protected devices}, should validate codes (e.g., software, configuration, or control commands) with firewalls before executing or relaying them. Hence, neither the firewalls nor the protected devices participate in malware diffusion. Consequently, the secured zones enforced by the firewalls will split the network into protected (i.e., green) and susceptible (gray) IoT/CPS devices. The design objective for spatial firewalls is to enforce sufficient secured zones, denoted as {\em critical percentage}, that ensures quarantined epidemics. Regardless of the infection/treatment rates, the firewalls provides a ubiquitous defense mechanism to thwart wide-spread diffusion of malware and  initiate informed (i.e., status-aware) software patching for infected devices.}

Interpreting the firewall design objective to the notion of percolation theory, it is required to remove the minimum number of vertices from the IoT/CPS network graph such that the largest connected subgraph does not percolate. In other words, it is required to place sufficient number of firewalls at effective locations such that the network formed by devices outside the secured zones operate in the subcritical regime. Otherwise, the epidemic spatial diffusion may get out of control and span the horizon. It is worth noting that the wireless network topology is dynamic due to devices mobility. Furthermore, the security defense mechanism should be general in case there is no prior knowledge about the underlying devices locations. {Hence, the spatial firewalls should be selected among the set of vertices $V$ in a RRG $G$, which accounts for devices wireless communication range as well as the random spatial locations of devices.}

A well established result in the literature for random graphs is that network percolation exhibit a phase transition phenomenon between the supercritical and subcritical regimes. This means that it is  feasible to quarantine epidemics in LSWN if the correct percentage of devices are protected. {Such critical percentage could be mathematically quantified via percolation models. However, different from conventional percolation problems, where individual vertices are randomly removed, in the spatial firewall problem a chunk of proximate vertices (i.e., all devices in the spatial secured zones including the firewalls, as well as all of their associated edges) are removed (c.f. Fig.~\ref{fig:SF1}). Hence, advanced percolation models, such as percolation on networks with holes~\cite{holes}, are required for the design, implementation, and assessment of spatial firewalls.}

Inspired by results from percolation theory, the firewalls design objective is decomposed to 1) find {\em minimum percentage} of spatial firewalls that guarantees quarantined malware,\footnote{Due to the involved CAPEX and OPEX of the spatial firewalls (e.g, hardware upgrade and software license), a cost efficient selection strategy should quarantine an epidemic with less critical intensity of firewalls.} 2) select the spatial locations of firewalls that minimizes the  average sizes of connected clusters of susceptible devices.  {This article examines four heuristic firewall selection schemes, namely, i) {\em random}, ii) {\em degree-aware}, iii) {\em random with minimum distance constraint (DC)}, and iv) {\em degree-aware with minimum DC}. Degree awareness is crucial because devices with high degree would participate more to an epidemic diffusion. Intuitively, devices with higher degrees are expected to have a significant impact on epidemic diffusion if selected as firewalls due to the high number of protected neighbors within the secured zone. The minimum DC among the firewalls guarantees good spatial distribution of secured zones across the network to thwart emerging epidemics in different locations.} 

A pictorial illustration of the firewalls is shown in Fig.~\ref{fig:SF1} for different selection strategies. While the same number of firewalls is selected in all cases, Figs.~\ref{fig:random_fire}, \ref{fig:degree_fire}, and \ref{fig:RDC_fire}, show poor spatial distributions of secured zones that fails to eliminate the giant connected component. Thus, a wide spread of the malware epidemic is viable. However, the degree-aware with minimum DC firewalls in Fig.~\ref{fig:degree_DC_fire} ensures a good spatial distribution of the secured zones and  succeeds to split the network into disjoint clusters. Hence, regardless of the infection/treatment rates, a malware injected to any susceptible device within a cluster is prevented to diffuse to other clusters in the network. {Upon the detection of the malware, firewalls inform the security administration to initiate localized software attestation/patching to all devices in the infected cluster.}


\subsection{Proof of Concept}
{While the proof-of-concept can be done analytically using tools from stochastic geometry and percolation theory, we resort to comprehensive Monte Carlo simulations to alleviate unnecessary mathematical details. Each simulation run realizes $4 \times 4$ km$^2$ random network  with wrap around boundaries. The devices are distributed in the simulation area according to a Poisson point process with intensity $\lambda=80$ device/km$^2$ (i.e., an average of 1280 device per simulation run). Each device is assumed to have a communication range of 200 m, where a connection is realized between two devices that lie within the communication range of each other. The firewalls are chosen among the network devices according to the implemented selection scheme, where each firewall enforces a circular secured zone with a radius of 200 m. The devices that fall within the secured zone are excluded and the global network connectivity is realized. Percolation is declared if a giant component of susceptible devices that spans the simulation area vertically and horizontally exists. Percolation means that the selected spatial firewalls fail to quarantine the malware infection. This is because a malware infiltrated to any of the devices within the giant component can create an epidemic outbreak within that component of susceptible devices without being obstructed by the selected spatial firewalls.} Furthermore, the number of disjoint clusters and the number of devices per each cluster are recorded. A pictorial illustration of a single simulation run is depicted in Fig.~\ref{fig:SF1}, where percolation is declared for the scenarios in Figs.~\ref{fig:random_fire}, \ref{fig:degree_fire}, \ref{fig:RDC_fire}.
\begin{figure}[t!]
	\centering
	\includegraphics[width=0.6\textwidth]{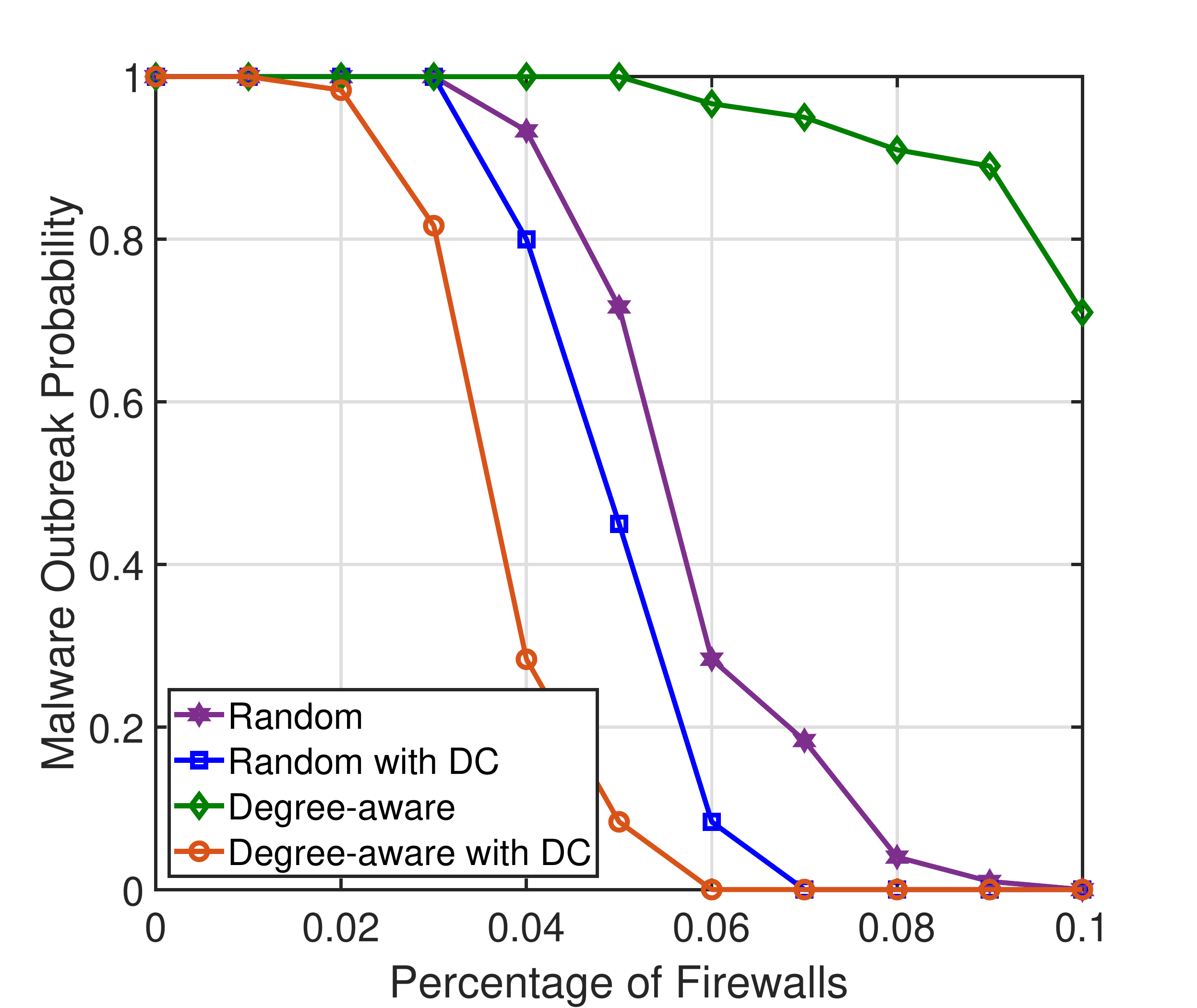}
	\caption{Probability of epidemic outbreak versus the percentage of firewalls.}
	\label{fig:Percolation}
\end{figure}

{Fig.~\ref{fig:Percolation} shows the malware outbreak probability versus the percentage of devices that are selected as firewalls. Surprisingly, random firewall selection significantly outperforms the degree-aware selection scheme. Such counter-intuitive behavior is due to the spatially correlated degree of devices. Hence, stand-alone degree awareness fails to quarantine the epidemic due to the spatial concentration of firewalls within small geographical regions, which leads to overlapped secured zones (see Fig.~\ref{fig:degree_fire}).  The random selection of firewalls provides a better spatial coverage of secured zones, and hence, succeeds to quarantine epidemics when the firewalls are sufficiently dense (i.e, critical percentage $9\%$). Incorporating minimum DC improves the spatial distribution of firewalls and reduces the required density of firewalls. The superior performance of degree-awareness with minimum DC  (critical percentage $6\%$) is due to the strategic selection of firewalls in terms of both spatial distribution of secured zones and number of devices covered by each secured zone.}

\begin{figure}[t!]
    \centering
    \includegraphics[width=0.6\textwidth]{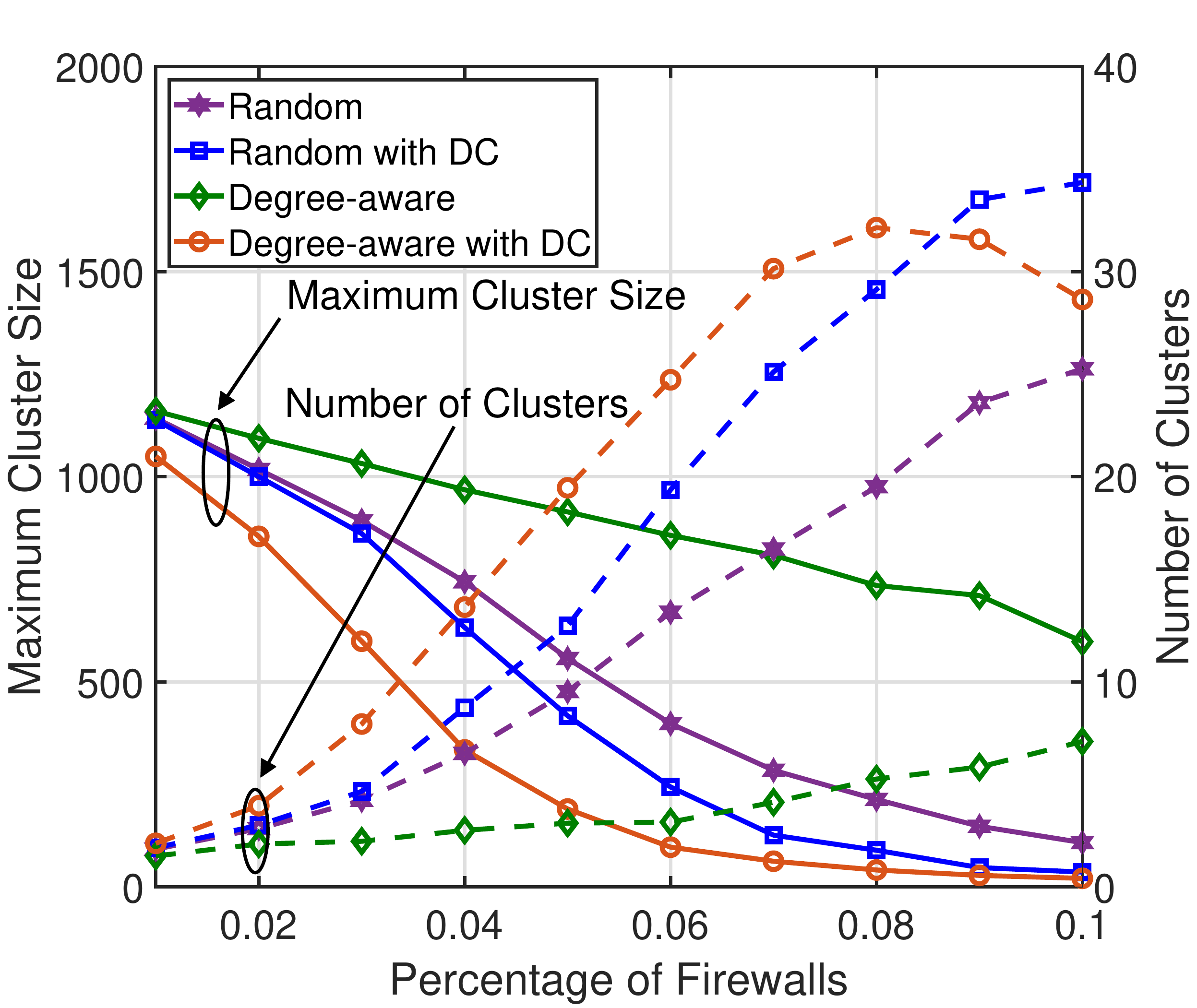}
	\caption{Number of disjoint clusters and size of the maximum cluster versus the percentage of firewalls.}
	\label{fig:Custersize}
\end{figure}

{Fig.~\ref{fig:Custersize} shows the number of susceptible clusters along with the number of devices in the largest cluster.  As the intensity of firewalls increases, large network components break into smaller susceptible clusters with less devices in each cluster. Hence, the number of clusters increases and the number of devices within each cluster decreases. Fig.~\ref{fig:Custersize} confirms the insights of Fig.~\ref{fig:Percolation} regarding the firewalls selection schemes. Degree and location awareness ensure that malware can only infect less number of devices within smaller geographical regions when compared to other sections schemes. It is worth noting that the number of clusters decreases in Fig.~\ref{fig:Custersize} when secured zones start to span and cover the entire simulation area, which leaves small spatial gaps to form clusters. }

\section{Acknowledgment}
Fig.~\ref{fig:SF0} was created by Heno Hwang, scientific illustrator at King Abdullah University of Science and Technology (KAUST).

\section{Conclusions}

This article overviews the vulnerabilities and cybersecurity threats in large-scale wireless IoT and CPS. The article then proposes a novel technique, denoted as spatial firewalls, to quarantine and cure malware epidemics in such IoT and CPS. In particular, we show that strategically selecting less than 10$\%$ of devices and equipping them with up-to-date anti-malware programs is sufficient to thwart malware epidemics. To this end, guidelines to design and characterize the impact of spatial firewalls is presented. Proof-of-concept numerical results are presented and several firewalls selection schemes are evaluated, namely, random and degree-aware with and without minimum distance constraints. Surprisingly, the random firewall selection outperforms degree-aware firewall selection, which is due to the spatially correlated degree of devices that lead to poor spatial distribution of firewalls. Adding a minimum distance constraint to degree-aware selection scheme significantly enhances its impact in terms of percentile of firewalls required to spatially quarantine malware epidemics and the size of infected clusters.

\bibliographystyle{IEEEtran}
\bibliography{IEEEabrv,references}
\section*{Biographies}
\textbf{Hesham Elsawy} [S'10, M'14, SM'17] received the Ph.D. degree in electrical engineering from the University of Manitoba, Canada, in 2014. He was a Post-Doctoral Fellow at the King Abdullah University of Science and Technology (KAUST), Saudi Arabia, a Research Assistant at TRTech, Winnipeg, MB, Canada, and a Telecommunication Engineer at the National Telecommunication Institute, Egypt. He is currently an Assistant Professor with the King Fahd University of Petroleum and Minerals (KFUPM), Saudi Arabia. His research interests include statistical modeling of wireless networks, stochastic geometry, and queueing analysis for wireless communication networks. He received several academic awards at the University of Manitoba, including the NSERC Industrial Postgraduate Scholarship (2010–2013), and the TRTech Graduate Students Fellowship (2010–2014). He has coauthored four award-winning papers that are recognized by the IEEE COMSOC Best Tutorial Paper Award, IEEE COMSOC Best Survey Paper Award, the Best Scientific Contribution Award to the IEEE International Symposium on Wireless Systems 2017, and the Best Paper Award in Small Cell and 5G Networks (SmallNets) Workshop of the 2015 IEEE International Conference on Communications (ICC). He is a recipient of the IEEE ComSoc Outstanding Young Researcher Award for Europe, Middle East, \& Africa Region in 2018. He is recognized as an exemplary reviewer by the IEEE Transactions on Communications (2014–2016), the IEEE Transactions on Wireless Communications in 2017 and 2018, and the IEEE Wireless Communications Letters in 2018.

\textbf{Mustafa A. Kishk} [S'16, M'18] is a postdoctoral research fellow in the communication theory lab at King Abdullah University of Science and Technology (KAUST). He received his B.Sc. and M.Sc. degree from Cairo University in 2013 and 2015, respectively, and his Ph.D. degree from Virginia Tech in 2018. His current research interests include stochastic geometry, energy harvesting wireless networks, UAV-enabled communication systems, and satellite communications.

\textbf{Mohamed-Slim Alouini} [S'94, M'98, SM'03, F'09] was
born in Tunis, Tunisia. He received the Ph.D. degree in Electrical Engineering
from the California Institute of Technology (Caltech), Pasadena,
CA, USA, in 1998. He served as a faculty member in the University of Minnesota,
Minneapolis, MN, USA, then in the Texas A\&M University at Qatar,
Education City, Doha, Qatar before joining King Abdullah University of
Science and Technology (KAUST), Thuwal, Makkah Province, Saudi
Arabia as a Professor of Electrical Engineering in 2009. His current
research interests include the modeling, design, and
performance analysis of wireless communication systems.


\end{document}